\documentclass[aps,prb,showpacs,preprint,amssymb,amsmath,floatfix]{revtex4-1}
\usepackage{graphicx}
\begin{document}
\title{Time-dependent embedding: surface electron emission}
\author{J. E. Inglesfield}
\affiliation{School of Physics and Astronomy,Cardiff University, 
The Parade, Cardiff, CF24 3AA, United Kingdom}
\email{JE.Inglesfield@googlemail.com}
\date{\today}
\begin{abstract}
An embedding method for solving the time-dependent Schr\"odinger equation is developed 
using the Dirac-Frenkel variational principle. Embedding allows the time-evolution of the
wavefunction to be calculated explicitly in a limited region of space, the region of physical 
interest, the embedding potential ensuring that the wavefunction satisfies the correct boundary 
conditions for matching on to the rest of the system. This is applied to a study of the excitation
of electrons at a metal surface, represented by a one-dimensional model potential for Cu(111). Time-dependent embedding potentials
are derived for replacing the bulk substrate, and the image potential and vacuum region
outside the surface, so that the calculation of electron excitation by a surface perturbation can 
be restricted to the surface itself. The excitation of the Shockley surface state and a continuum 
bulk state is studied, and the time-structure of the resulting currents analysed.
Non-linear effects and the time taken for the current to arrive outside the surface are discussed.
The method shows a clear distinction between emission from the localized surface state, where the charge is steadily depleted, and the extended continuum
state where the current emitted into the vacuum is compensated by current approaching the 
surface from the bulk.
\end{abstract}
\maketitle
\section{Introduction} \label{sec:intro}
The question ``How long does it take?" arises in many areas of condensed matter physics, 
in photoemission from surfaces \cite{Cavalieri:2007fk}, electron tunnelling through barriers 
\cite{RevModPhys.66.217,Carvalho:2002kx}, relaxation after the creation of a core hole \cite{PhysRevLett.101.046101} 
and so on. In this paper we shall study the time-dependence of electron emission 
from a metal surface, using an embedding formalism which we have recently developed
\cite{Inglesfield:2008ys} and now further improved. 

The development of attosecond streaking spectroscopy, using ultra-short laser pulses, has raised 
interesting questions about the time it takes for electrons to be emitted in photoemission \cite{Cavalieri:2007fk}, 
and recent computational studies have tackled this problem \cite{PhysRevLett.102.177401,PhysRevB.81.075440}. 
In these time-dependent calculations, the surface and substrate are represented by a slab with a finite 
region of vacuum outside, and inevitably an electron excited by the time-dependent
perturbation reaches the edge. How to deal with this is a familiar problem
in time-dependent calculations in atomic physics, and it is often handled (as in the calculations
just mentioned) by introducing an absorbing potential at the boundaries \cite{PhysRevLett.66.2601}. It is important to
eliminate spurious back reflections, and exact boundary potentials have been developed 
which ensure that the wavefunction has the correct boundary conditions to match on to 
the rest of space \cite{PhysRevB.49.2904,Ehrhardt:1999vn}. 
This is in fact the idea of embedding \cite{Inglesfield:1981vn,Inglesfield:2008ys}: 
we solve the Schr\"odinger equation in the region of space of interest, which we call region I,
with an embedding potential at the boundary of the region to ensure that the wavefunction matches correctly on to the external region, region II. 

The original time-independent embedding method \cite{Inglesfield:1981vn} has proved
useful for calculations of surface electronic structure \cite{Benesh:1984ys} -- only the surface 
region, the top few atomic layers plus the near-surface vacuum region 
are treated explicitly, the bulk crystal and the semi-infinite vacuum being replaced by embedding 
potentials derived from the substrate Green functions. The embedding method treats the bulk
continuum states at the surface, as well as localized surface states correctly, and gives a very
accurate description of surface electronic structure \cite{Ishida:2001sh}. Our time-dependent
embedding method builds on time-independent embedding and in the previous paper 
\cite{Inglesfield:2008ys}  we transformed the embedding potential as a function of energy
into a function of time, and developed an embedded time-dependent Schr\"odinger equation
for region I.

This time-dependent embedding used the results of earlier papers in which the problem of the
exact termination of the spatially discretized Schr\"odinger equation was studied
\cite{PhysRevA.56.763,Ehrhardt:1999vn}. However, in our method, we use a basis set expansion
of the wavefunction in region I, with time-varying coefficients, to solve the Schr\"odinger
equation. Recent time-dependent calculations on electron transport through molecules have
used self-energies to describe the coupling of the molecular wavefunctions
to the metal electrodes \cite{Mujica:2003wb}. The self-energy is in fact an embedding potential in 
a tight-binding representation \cite{Baraff:1986le}, and this approach is related to our
time-dependent embedding.

We have made several advances in the time-dependent formalism since the earlier paper \cite{Inglesfield:2008ys}, 
and we present these here, together with results on the emission of electrons from bulk and
surface states at the Cu(111) surface. In section \ref{sec:formalism} we shall give a new derivation 
of the method based on the Dirac-Frenkel variational principle \cite{Dirac:1930uq,Frenkel:1934fk}, appropriate to 
solutions of the time-dependent Schr\"odinger equation. This avoids some of the weaknesses of 
our earlier derivation, which was based partly on analogy with time-independent embedding. In 
our application to emission of electrons from the Cu(111) surface, we use a one-dimensional 
model potential due to Chulkov \emph{et al.} \cite{Chulkov:1997pr,Chulkov:1999fh} 
to describe the bulk, the surface, and the image 
potential outside the surface -- this gives a good description of the electronic structure around 
the s-p band gap, with the Shockley and image potential-induced surface states, and has been
widely used in surface calculations \cite{Echenique:2004bd,PhysRevLett.102.177401}. 
In section \ref{sec:surfaceemb} we shall calculate the
time-dependent embedding potentials for replacing the semi-infinite bulk and vacuum
regions, making use of some recent results for the time-independent embedding potentials 
appropriate to the Chulkov surface potential \cite{Chulkov:1997pr,Chulkov:1999fh}. Tests of the method, in which we apply a time-dependent 
surface perturbation, are described in section \ref{sec:testing}. A crucial test is provided by the 
continuity equation -- the number of electrons in the surface region plus the time-integrated currents 
across the boundaries of the embedded region should be constant. This is satisfied to a high 
degree of accuracy. 

In section \ref{sec:results} we give results for the electron current through the boundaries of 
the embedded surface region, into the bulk and into the vacuum, when a time-dependent 
perturbing potential is applied at the surface. We find that after initial transients, the current
into the vacuum settles down, with an average value given by 
Fermi's Golden Rule \cite{Merzbacher:1998fk}. 
With a low frequency of surface perturbation, non-linear oscillations appear in the 
current, due to the interference between first order and second order processes.
We also study the time it takes for the electrons to arrive at various distances from the surface, 
comparing the arrival time with the classical value, and for the time for the current to peak.
Our most surprising results concern the comparison of currents into the bulk and into the vacuum.
In the case of emission from a bulk continuum state, the current leaving the surface region into 
the vacuum is balanced by the current entering the surface from the bulk, so that the
charge in the surface region stays constant on average; however, in emission from a surface 
state, currents of very similar magnitude leave the surface in both directions.
Finally, in section \ref{sec:outlook} we give a brief outlook for the method.

Atomic units are used throughout this paper, with $e=\hbar=m_e=1$. The atomic unit of time
$=2.418884\times10^{-17}$ s, so that 1 fs = 41.34138 a.u.

\section{Time-dependent embedding formalism} \label{sec:formalism}
\begin{figure}[h]
\begin{center}
\includegraphics[width=10cm] {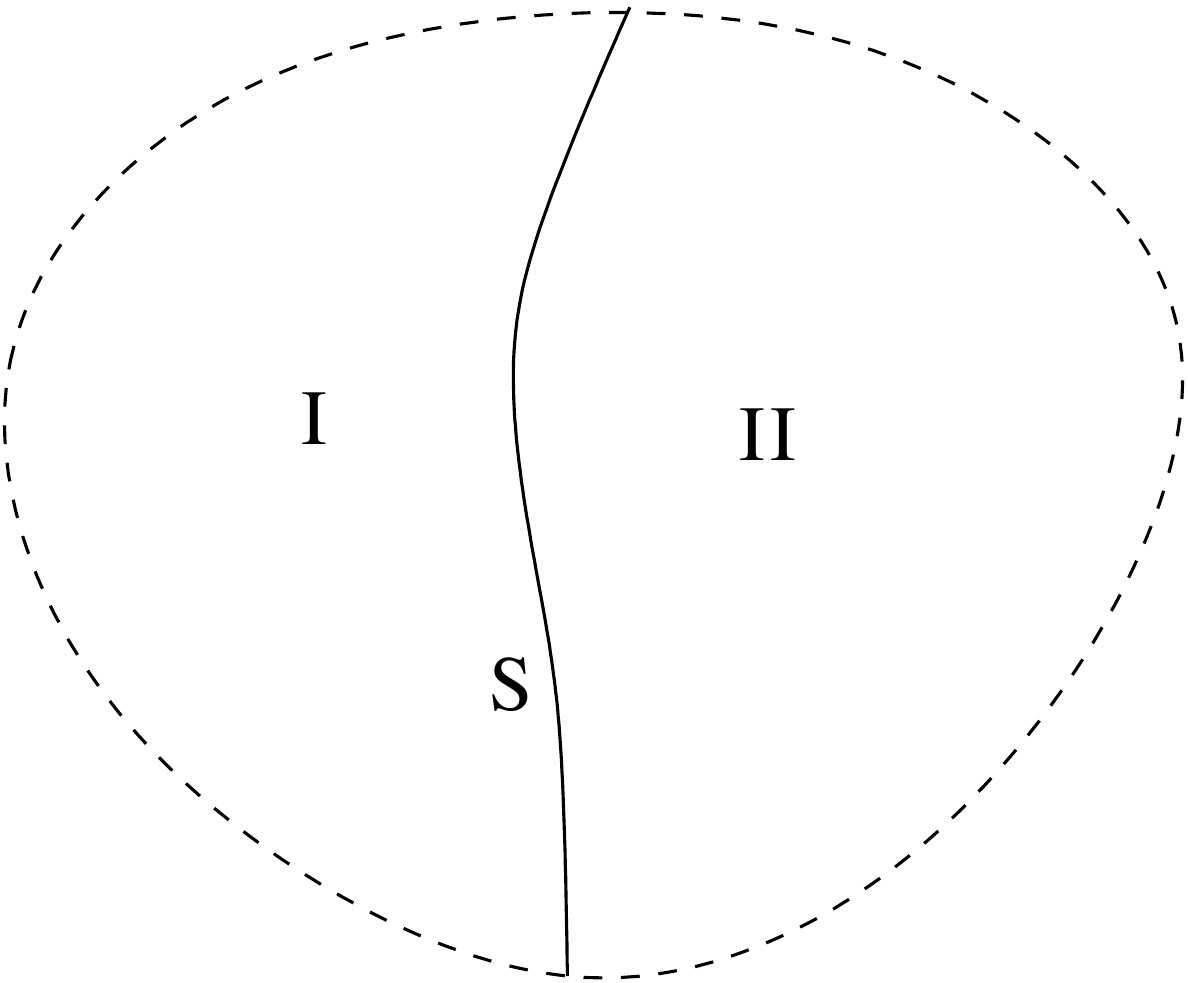}
\caption{Region I is embedded on to region II over surface S. The Schr\"odinger equation is solved explicitly in region I,
region II being replaced by the embedding potential over S.} \label{fig:regions}
\end{center}
\end{figure}

The idea of embedding is that we solve the Schr\"odinger equation explicitly only in region I,
the rest of the system, which we call region II, being replaced by an embedding potential over the interface S (figure \ref{fig:regions}). 
We developed the embedding method for the time-independent
Schr\"odinger equation using a variational principle \cite{Inglesfield:1981vn}, 
and showed that the wavefunction satisfies the following equation in
region I (we include S within region I),
\begin{eqnarray}
\lefteqn{\left(-\frac{1}{2}\nabla^2+V(\mathbf r)\right)\phi(\mathbf{r})+\delta(\mathbf{r}-\mathbf{r}_\mathrm{S})\left[\frac{1}{2}
\frac{\partial\phi}{\partial n_\mathrm{S}}\right.}    \nonumber\\
&&\hspace{2cm}+\left.\int_S d\mathbf{r}_S'\left(G_0^{-1}(\mathbf{r}_{\mathrm{S}},\mathbf{r}_{\mathrm{S}}';\epsilon)+
(E-\epsilon)\frac{\partial G_0^{-1}}{\partial \epsilon}\right)
\phi(\mathbf{r}'_S)\right]=E\phi(\mathbf{r}).
\label{eq:en_emb}
\end{eqnarray}
$G^{-1}_0(\mathbf{r}_{\mathrm{S}},\mathbf{r}_{\mathrm{S}}';\epsilon)$ is the embedding potential, 
a generalized logarithmic derivative of the solution of the Schr\"odinger equation in region II at energy $\epsilon$,
\begin{equation}
\frac{\partial\psi(\mathbf{r}_{\mathrm{S}})}{\partial n_{\mathrm{S}}}=-2\int_{\mathrm{S}} d\mathbf{r}'_{\mathrm{S}}
G^{-1}_0(\mathbf{r}_{\mathrm{S}},\mathbf{r}_{\mathrm{S}}';\epsilon)\psi(\mathbf{r}'_{\mathrm{S}}),
\label{eq:en_logderiv}
\end{equation}
where we take the normal derivative on S outwards from region I into II.
The two terms in the round brackets in (\ref{eq:en_emb}), the embedding potential evaluated at energy parameter $\epsilon$ 
and the energy derivative term, give $G_0^{-1}$ at energy $E$, the energy at which we are solving the Schr\"odinger equation,
to first order in $(E-\epsilon)$. 
The surface terms inside the square brackets then vanish when $\phi$ matches in amplitude and derivative on to
the solution of the Schr\"odinger equation in region II, giving the correctly embedded solution of the Schr\"odinger equation.

We now turn to the time-dependent embedding problem, where we shall assume that the time-dependent potential, such as
an external perturbation, is confined to region I. First we need a relationship analogous to (\ref{eq:en_logderiv}) for the 
solution of the time-dependent Schr\"odinger equation in II. 
This is given by \cite{PhysRevA.56.763,Ehrhardt:1999vn}
\begin{equation}
  \frac{\partial\psi(\mathbf{r}_{\mathrm{S}},t)}{\partial
    n_{\mathrm{S}}}=-2\int_{\mathrm{S}}d\mathbf{r}'_{\mathrm{S}}
  \int^t_{-\infty}dt'
  \bar{G}^{-1}_0(\mathbf{r}_{\mathrm{S}},\mathbf{r}'_{\mathrm{S}};t-t')
  \frac{\partial\psi(\mathbf{r}'_{\mathrm{S}},t')}{\partial t'},
\label{eq:dir_neu}
\end{equation}
where the time-dependent embedding potential is given by,
\begin{equation}
\bar{G}^{-1}_0(\mathbf{r}_{\mathrm{S}},\mathbf{r}'_{\mathrm{S}};t)=\frac{1}{2\pi}\int_{-\infty}^{+\infty} d\epsilon
\exp(-\mathrm{i}\epsilon t)\frac{G^{-1}_0(\mathbf{r}_{\mathrm{S}},\mathbf{r}'_{\mathrm{S}};\epsilon)}{-\mathrm{i}\epsilon}.
\label{eq:mod_FT}
\end{equation}
Equation (\ref{eq:dir_neu}) is almost the Fourier transform of (\ref{eq:en_logderiv}), but it contains 
the time-derivative $\partial\psi/\partial t'$; this compensates for the convergence factor of $-1/\mathrm{i}\epsilon$ in the
transform (\ref{eq:mod_FT}) of the embedding potential. 
When we evaluate this transform, we take $\epsilon$ to lie just above the real axis, 
above the singularities in the Green function, so that $\bar{G}^{-1}_0(t)=0$ for $t<0$.

With the time-dependent version of the generalized logarithmic derivative (\ref{eq:dir_neu}), we can now derive the time-dependent embedded
Schr\"odinger equation, and unlike our original derivation \cite{Inglesfield:2008ys}, here we shall use a variational method. 
There are, in fact, several stationary principles for the time-dependent Schr\"odinger
equation \cite{McLachlan:1964uq}, and we choose the Dirac-Frenkel variational principle \cite{Dirac:1930uq,Frenkel:1934fk}, which
gives the functional variation 
\begin{equation}
  \delta I=\int \!
  d\mathbf{r}\delta\Psi^\ast(\mathbf{r},t)\left[H-\mathrm{i}
    \frac{\partial}
    {\partial t}\right]\Psi(\mathbf{r},t)=0, \label{eq:var1}
\end{equation} 
for small variations $\delta\Psi^\ast$ of the time-dependent wavefunction. The Hamiltonian $H$  in (\ref{eq:var1}) is given by
\begin{equation}
H=-\frac{1}{2}\nabla^2+V(\mathbf{r},t),
\label{eq:ham}
\end{equation}
and the integral is over the combined regions I + II. 

We use the same form of
wavefunction $\Psi$ as in the original 
embedding method \cite{Inglesfield:1981vn} -- in region I, $\Psi$ is given by a trial wavefunction
$\phi(\mathbf{r},t)$, and in region II by the exact solution of the
time-dependent Schr\"odinger equation $\psi(\mathbf{r},t)$, which
matches in amplitude on to $\phi(\mathbf{r}_{\mathrm{S}},t)$ over the boundary S between the two
regions ($\phi$ and $\psi$ are assumed to be 
zero over the outer boundary of the combined region I + II, figure \ref{fig:regions}).
Evaluating the integrand in (\ref{eq:var1}) with this form of $\Psi$, the volume
integral splits into the two regions, with an additional contribution on S coming 
from the kinetic energy operator $-\frac{1}{2}\nabla^2$  acting on the 
discontinuity in derivative between $\phi$ and $\psi$,
\begin{equation}
\delta I=\int_{\mathrm{I}} d\mathbf{r}\delta\phi^\ast\left[H\phi-\mathrm{i}\frac{\partial\phi}{\partial t}\right]
+\int_{\mathrm{II}} d\mathbf{r}\delta\psi^\ast\left[H\psi-\mathrm{i}\frac{\partial\psi}{\partial t}\right]
+\frac{1}{2}\int_{\mathrm{S}}d\mathbf{r}_{\mathrm{S}}\delta\phi^\ast\left[\frac{\partial\phi}{\partial n_{\mathrm{S}}}
-\frac{\partial\psi}{\partial n_{\mathrm{S}}}\right]. 
\label{eq:var2}
\end{equation}
But as $\psi$ is an exact solution of the time-dependent Schr\"odinger equation in region II,
the second integral is zero. Also using (\ref{eq:dir_neu}) and the constraint that 
$\psi(\mathbf{r}_{\mathrm{S}},t)=\phi(\mathbf{r}_{\mathrm{S}},t)$, the term involving $\partial\psi/\partial n_{\mathrm{S}}$
can be rewritten as
\begin{equation}
-\frac{1}{2}\int_{\mathrm{S}}d\mathbf{r}_{\mathrm{S}}\delta\phi^\ast\frac{\partial\psi}{\partial n_{\mathrm{S}}}=
\int_{\mathrm{S}}d\mathbf{r}_{\mathrm{S}}\int_{\mathrm{S}}d\mathbf{r}'_{\mathrm{S}}\int_{-\infty}^t
\delta\phi^\ast(\mathbf{r}_{\mathrm{S}},t)\bar{G}^{-1}_0(\mathbf{r}_{\mathrm{S}},\mathbf{r}'_{\mathrm{S}};t-t')
\frac{\partial\phi(\mathbf{r}'_{\mathrm{S}},t')}{\partial t'}.
\label{eq:embpot3}
\end{equation}
This assumes that $\phi$ on S is zero in the distant past, and $\psi$ itself is zero -- in other words,
$\psi$ grows into region II from region I as time proceeds.
The variation of the functional then becomes an expression only involving $\phi$ in region I and on S, with the
embedding potential accounting for region II,
\begin{eqnarray}
\delta I&=&\int_{\mathrm{I}} d\mathbf{r}\delta\phi^\ast(\mathbf{r},t)\left[H-\mathrm{i}\frac{\partial}{\partial t}\right]\phi(\mathbf{r},t)
+\frac{1}{2}\int_{\mathrm{S}}d\mathbf{r}_{\mathrm{S}}\delta\phi^\ast(\mathbf{r}_{\mathrm{S}},t) \frac{\partial\phi(\mathbf{r}_{\mathrm{S}},t)}
{\partial n_{\mathrm{S}}}
\nonumber\\
&+&\int_{\mathrm{S}}d\mathbf{r}_{\mathrm{S}}\int_{\mathrm{S}}d\mathbf{r}'_{\mathrm{S}}\int_{-\infty}^t
\delta\phi^\ast(\mathbf{r}_{\mathrm{S}},t)\bar{G}^{-1}_0(\mathbf{r}_{\mathrm{S}},\mathbf{r}'_{\mathrm{S}};t-t')
\frac{\partial\phi(\mathbf{r}'_{\mathrm{S}},t')}{\partial t'}. \label{eq:var3}
\end{eqnarray}

We see that $\delta I=0$ for arbitrary variations of $\delta\phi^\ast$ if $\phi$ satisfies the time-dependent Schr\"odinger
equation in region I analogous to (\ref{eq:en_emb}),
\begin{eqnarray}
\lefteqn{\left(-\frac{1}{2}\nabla^2+V({\mathbf r},t)\right)\phi(\mathbf{r},t)+\delta(\mathbf{r}-\mathbf{r}_S)\left[\frac{1}{2}
\frac{\partial\phi}{\partial n_S}\right.}    \nonumber\\
&&\hspace{3cm}+\left.\int_S d\mathbf{r}_S'\int_{-\infty}^t dt' \bar{G}_0^{-1}(\mathbf{r}_{\mathrm{S}},\mathbf{r}_{\mathrm{S}}';t-t')
\frac{\partial\phi(\mathbf{r}'_S,t')}{\partial t'}\right]=\mathrm{i}\frac{\partial\phi}{\partial t}.
\label{eq:time_emb}
\end{eqnarray}
We can also
use (\ref{eq:var3}) to obtain the time-dependent matrix equation, with which we solve the embedded time-dependent
Schr\"odinger equation in practice. Let us expand the trial function $\phi$ in terms of a set of basis functions $\chi_i$,
\begin{equation}
\phi(\mathbf{r},t)=\sum_i a_i(t)\chi_i(\mathbf{r})   
\label{eq:exp}
\end{equation}
-- it is convenient both in the formalism and in coding if the basis functions are orthonormal when 
integrated over region I. Substituting this expansion into (\ref{eq:var3}) gives
\begin{equation}
\delta I=\sum_{ij}\delta a^\ast_i(t) \bar{H}_{ij}(t)a_j(t)-\mathrm{i}\sum_i\delta a^\ast_i(t)\frac{da_i}{dt}+\sum_{ij}\delta a^\ast_i(t)
\int_{-\infty}^t dt'\bar{\Sigma}_{ij}(t-t')\frac{da_j}{dt'},   \label{eq:var4}
\end{equation}
where the Hamiltonian matrix $\bar{H}_{ij}$ includes the surface derivative term in (\ref{eq:var3}),
\begin{equation}
\bar{H}_{ij}(t)=\frac{1}{2}\int_{\mathrm{I}} d\mathbf{r}\nabla\chi_i(\mathbf{r})\cdot\nabla\chi_j(\mathbf{r})
+\int_{\mathrm{I}} d\mathbf{r}\chi_i(\mathbf{r})V(\mathbf{r},t)\chi_j(\mathbf{r}),    \label{eq:hammat1}
\end{equation}
and the embedding matrix $\bar{\Sigma}_{ij}$ is given by
\begin{equation}
\bar{\Sigma}_{ij}(t)=\int_{\mathrm{S}}d\mathbf{r}_{\mathrm{S}}\int_{\mathrm{S}}d\mathbf{r}'_{\mathrm{S}}
\chi_i(\mathbf{r}_{\mathrm{S}})\bar{G}^{-1}_0(\mathbf{r}_{\mathrm{S}},\mathbf{r}'_{\mathrm{S}};t)\chi_j(\mathbf{r}_{\mathrm{S}}).
\label{eq:embmat}
\end{equation}
Then $\delta I=0$ for arbitrary $\delta a^\ast_i(t)$, and the variational principle is satisfied, when
\begin{equation}
\sum_j\left[\bar{H}_{ij}(t)a_j(t)+\int_{-\infty}^t dt'\bar{\Sigma}_{ij}(t-t')\frac{da_j}{dt'}\right]=\mathrm{i}\frac{da_i}{dt}
\label{eq:matvar1}
\end{equation}
-- the time-dependent embedded Schr\"odinger equation in matrix form. This equation was derived in our earlier
paper \cite{Inglesfield:2008ys} in a less straightforward way.

This formalism, which handles the evolution of a state initially localized in region I, needs modifying to study the 
time evolution of bulk continuum states at the surface, as well as the discrete surface states. As in earlier surface embedding work \cite{Inglesfield:2001il},
we take region I to be the surface region, that is, the top few atomic layers plus the adjacent image potential + vacuum region, 
and region II to be the rest of the system, the semi-infinite crystal substrate on one side, and the semi-infinite 
vacuum on the other. The initial wavefunction then extends \emph{beyond} region I -- obviously so for 
continuum states, but even the exponentially decaying surface states extend some distance into the bulk and 
vacuum into region II. The time-evolution of extended states was also treated in our previous paper \cite{Inglesfield:2008ys}, 
but here we use the Dirac-Frenkel variational principle (\ref{eq:var1}) to develop the formalism
more consistently.

The unperturbed wavefunction $\Xi(\mathbf{r},t)$ is a stationary state of the Hamiltonian $H_0$, 
\begin{equation}
H_0=-\frac{1}{2}\nabla^2+V(\mathbf{r}), \label{eq:ind}
\end{equation}
where $V(\mathbf{r})$ is the time-independent potential in I + II. We have
\begin{equation}
\Xi(\mathbf{r},t)=\xi(\mathbf{r})\exp(-\mathrm{i}Et),\;\mbox{with}\;H_0\xi=E\xi.   \label{eq:eig}
\end{equation}
Let us now switch on the perturbing potential $\delta V(\mathbf{r},t)$
in region I at time $t=0$, and see how this wavefunction develops. 
As before we shall write the full time-dependent Hamiltonian as $H$, and the evolving wavefunction as $\Psi$, with
\begin{equation}
H=H_0,\;\;\;\Psi(\mathbf{r},t)=\Xi(\mathbf{r},t),\;\;\;t\le 0, \label{eq:tle}
\end{equation}
and for $t>0$,
\begin{eqnarray}
H=H_0+\delta V(\mathbf{r},t),\;\; \Psi(\mathbf{r},t)&=&\Xi(\mathbf{r},t)+\phi(\mathbf{r},t)\;\;\;\mathbf{r}\;\mbox{in region I}\nonumber\\
H=H_0,\;\;\Psi(\mathbf{r},t)&=&\Xi(\mathbf{r},t)+\psi(\mathbf{r},t)\;\;\;\mathbf{r}\;\mbox{in region II}. \label{eq:tgr}
\end{eqnarray}
Again, $\phi(\mathbf{r},t)$ is the trial function in region I, and $\psi(\mathbf{r},t)$ is the exact solution of the time-dependent
Schr\"odinger equation in region II, which matches on to $\phi(\mathbf{r}_{\mathrm{S}},t)$ over S. So $\Xi+\psi$ satisfies 
the time-dependent Schr\"odinger equation in region II, and the functional variation (\ref{eq:var1}) simplifies to
\begin{eqnarray}
\delta I&=&\int_{\mathrm{I}} d\mathbf{r}\delta\phi^\ast(\mathbf{r},t)\left[H\phi(\mathbf{r},t)+\delta V\Xi(\mathbf{r},t)
-\mathrm{i}\frac{\partial\phi(\mathbf{r},t)}{\partial t}\right]
+\frac{1}{2}\int_{\mathrm{S}}d\mathbf{r}_{\mathrm{S}}\delta\phi^\ast(\mathbf{r}_{\mathrm{S}},t) \frac{\partial\phi(\mathbf{r}_{\mathrm{S}},t)}
{\partial n_{\mathrm{S}}} \nonumber\\
&+&\int_{\mathrm{S}}d\mathbf{r}_{\mathrm{S}}\int_{\mathrm{S}}d\mathbf{r}'_{\mathrm{S}}\int_0^t
\delta\phi^\ast(\mathbf{r}_{\mathrm{S}},t)\bar{G}^{-1}_0(\mathbf{r}_{\mathrm{S}},\mathbf{r}'_{\mathrm{S}};t-t')
\frac{\partial\phi(\mathbf{r}'_{\mathrm{S}},t')}{\partial t'},\;\;\;t>0 \label{eq:var5}  
\end{eqnarray}
-- the same as (\ref{eq:var3}), apart from the extra inhomogeneous term $\delta V\Xi$ in the first integral, 
and in the time integral we have taken a lower limit of $t=0$, the time at which the perturbation is switched on.

We can finally write down the matrix equation for the time evolution of extended states.
Substituting expansion (\ref{eq:exp}) into (\ref{eq:var5}) gives
\begin{eqnarray}
\delta I&=&\sum_{ij}\delta a^\ast_i(t)\bar{H}_{ij}(t)a_j(t)+\sum_i \delta a^\ast_i(t)e_i(t)-\mathrm{i}\sum_i\delta a^\ast_i(t)\frac{da_i}{dt} 
       \nonumber\\
&+&\sum_{ij}\delta a^\ast_i(t)\int_0^t dt'\bar{\Sigma}_{ij}(t-t')\frac{da_j}{dt'},\;\;\;  t>0,   \label{eq:var6}
\end{eqnarray}
where the Hamiltonian matrix now includes the perturbing potential,
\begin{equation}
\bar{H}_{ij}(t)=\frac{1}{2}\int_{\mathrm{I}} d\mathbf{r}\nabla\chi_i(\mathbf{r})\cdot\nabla\chi_j(\mathbf{r})
+\int_{\mathrm{I}} d\mathbf{r}\chi_i(\mathbf{r})[V(\mathbf{r})+\delta V(\mathbf{r},t)]\chi_j(\mathbf{r}),    \label{eq:hammat2}
\end{equation}
and vector $e_i$ comes from the unperturbed state,
\begin{equation}
e_i(t)=\int_\mathrm{I}d\mathbf{r}\chi_i(\mathbf{r})\delta V(\mathbf{r},t)\Xi(\mathbf{r},t).   \label{eq:inhom1}
\end{equation}
The variational principle then gives the inhomogeneous time-dependent matrix equation,
\begin{equation}
\sum_j\left[\bar{H}_{ij}(t)a_j(t)+\int_0^t dt'\bar{\Sigma}_{ij}(t-t')\frac{da_j}{dt'}\right]+e_i(t)
=\mathrm{i}\frac{da_i}{dt}, \;\;\;t>0, \label{eq:matvar2}
\end{equation}
which we can solve for $a_i(t)$ with the initial condition that $a_i(t=0)=0$. Knowing the $a_i$ gives us $\phi$, hence
$\Psi$ -- the time-evolving wavefunction in region I, embedded on to the rest of the system.

\section{Surface embedding potentials} \label{sec:surfaceemb}
\begin{figure}[h]
\begin{center}
\includegraphics[width=14cm] {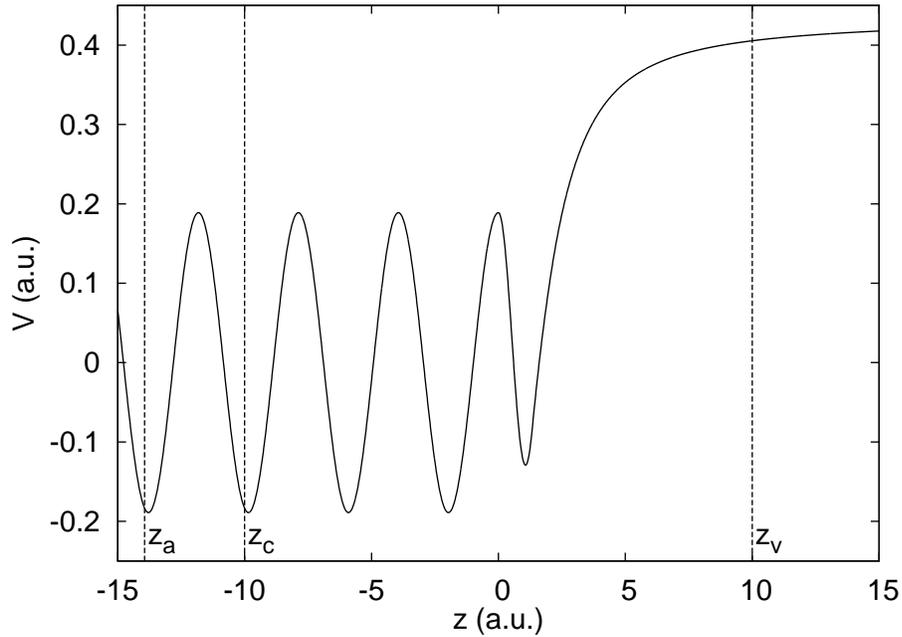}
\caption{One-dimensional potential for modelling the Cu(111)
surface. At $z_c$ the surface region is embedded on to the semi-infinite crystal potential, and at $z_v$  it is embedded on to
the vacuum Coulomb tail. The Schr\"odinger equation is integrated between $z_a$ and $z_c$ to find $G_c^{-1}(\epsilon)$.} \label{fig:cupot}
\end{center}
\end{figure}

We shall apply this formalism to study the excitation of electron states at a metal
surface under the influence of a time-dependent perturbing potential. In the previous paper \cite{Inglesfield:2008ys} 
we applied this formalism to a jellium surface, but
here we shall consider a more realistic model of the surface, one which can support surface states as well as the bulk
continuum. This is the one-dimensional model potential $V(z)$ developed by Chulkov \emph{et al.} \cite{Chulkov:1997pr,Chulkov:1999fh}, 
which has the form, 
\begin{equation}
V(z)=\left\{\begin{array}{ll}
A_1\cos(2\pi z/a), & z<0\\
-A_{10}-A_{20}+A_2\cos(\beta z), & 0<z<z_1\\
-A_{10}+A_3\exp[-\alpha(z-z_1)], & z_1<z<z_{\mathrm{im}}\\
-A_{10}+(\exp[-\lambda(z-z_{\mathrm{im}})]-1)/4(z-z_{\mathrm{im}}), & z>z_{\mathrm{im}}.
\end{array}\right.   \label{eq:chulkov}
\end{equation}
Here $a$ is the interlayer spacing, parameter $A_1$ reproduces the width of the bulk energy gap,  $A_{10}$ is the inner potential
(a difference with Chulkov \emph{et al.} \cite {Chulkov:1999fh} is that 
we take the average potential in the bulk as the energy zero, rather than the vacuum potential),
and $A_2$ and $\beta$ are fitted to surface state energies. 
The remaining parameters are found by the requirement that the
potential and its derivative are continuous across all the boundaries. 
The electrons are of course moving in three dimensions, but the motion parallel 
to the surface is free electron-like, and we shall only consider motion in the $z$-direction. 
Figure \ref{fig:cupot} shows the resulting surface potential for Cu(111), the surface we shall be studying in this paper, 
calculated with the parameters given in table \ref{table:chulkov}.
\begin{table}[t]
\begin{center}
\begin{tabular}{|c|l|l|l|l|} \hline
region with $z<0$ & $a=3.94$ & $A_1=0.18889$ & &\\ \hline
$0<z<z_1$ & $\beta=2.9416$ & $A_{10}=-0.43713$ & $A_2=0.15905$ & $A_{20}=0.40729$ \\ \hline
$z_1<z<z_{\mathrm{im}}$ & $\alpha=0.63650$ & $z_1=1.33499$ & $A_3=-0.51975$ &\\ \hline
$z>z_{\mathrm{im}}$ & $\lambda=1.27300$ & $z_{\mathrm{im}}=2.10562$ & & \\ \hline
\end{tabular}
\end{center}
\caption{Parameters used in the Chulkov potential for Cu(111) for different regions near the surface. 
$a$, $A_1$, $A_{10}$, $A_2$ and $\beta$ are tabulated by Chulkov \emph{et al.} \cite{Chulkov:1999fh}, 
and the other parameters are derived; all parameters are given in atomic units.}
\label{table:chulkov}
\end{table}

We now derive the time-dependent embedding potentials for embedding the surface region on to the substrates on either side 
(figure \ref{fig:cupot}), $\bar{G}_c^{-1}(t)$ for embedding on to the semi-infinite crystal substrate at $z_c$, and $\bar{G}_v^{-1}(t)$
for embedding on to the vacuum at $z_v$. 
We start with the embedding potential for replacing the semi-infinite crystal, first calculating the energy-dependent embedding potential
$G_c^{-1}(\epsilon)$ for subsequent Fourier transforming according to (\ref{eq:mod_FT}). As we have shown elsewhere \cite{Inglesfield:2010fk}, 
the first step is to integrate the Schr\"odinger equation at energy $\epsilon$ through a bulk unit cell in each direction to give us
two independent solutions, $\phi_1$ and $\phi_2$. Taking the unit cell to lie between $z_a$ and $z_c$ (figure \ref{fig:cupot}), the boundary conditions for determining $\phi_1$ and $\phi_2$ are as follows,
\begin{eqnarray}
\phi_1(z_a)=1,\;\phi'_1(z_a)=0,&\;\;\;&\mbox{integrate from }z_a\mbox{ to }z_c\nonumber\\
\phi_2(z_c)=1,\;\phi'_2(z_c)=0,&\;\;\;&\mbox{integrate from }z_c\mbox{ to }z_a. \label{eq:boundcond}
\end{eqnarray} 
The corresponding wave-vector $k$ of the bulk band structure
is then given by the remarkably simple expression,
\begin{equation}
\cos(ka)=\frac{\phi_1(z_c)+\phi_2(z_a)}{2}  \label{eq:bandstructure}
\end{equation}
-- a related result has been given by Butti \cite{Butti:2005fk}, and earlier by Kohn \cite{Kohn:1959if}. Using the property that it is 
a logarithmic derivative, it can be shown \cite{Inglesfield:2010fk} that the embedding potential for embedding on to the 
semi-infinite crystal to the left of $z_c$ is given by
\begin{equation}
G_c^{-1}(\epsilon)=\frac{W(\phi_1,\phi_2)}{2[\exp(-\mathrm{i}ka)-\phi_1(z_c)]},    \label{eq:crystenemb}
\end{equation}
where $W$ is the Wronskian. 
For causality, the Schr\"odinger equation is solved at $\epsilon$ with a positive imaginary part (which may be
infinitesimal), and $k$ is chosen to have negative imaginary part, corresponding to a wave travelling or decaying into the crystal to the left.
\begin{figure}[h]
\begin{center}
\includegraphics[width=14cm] {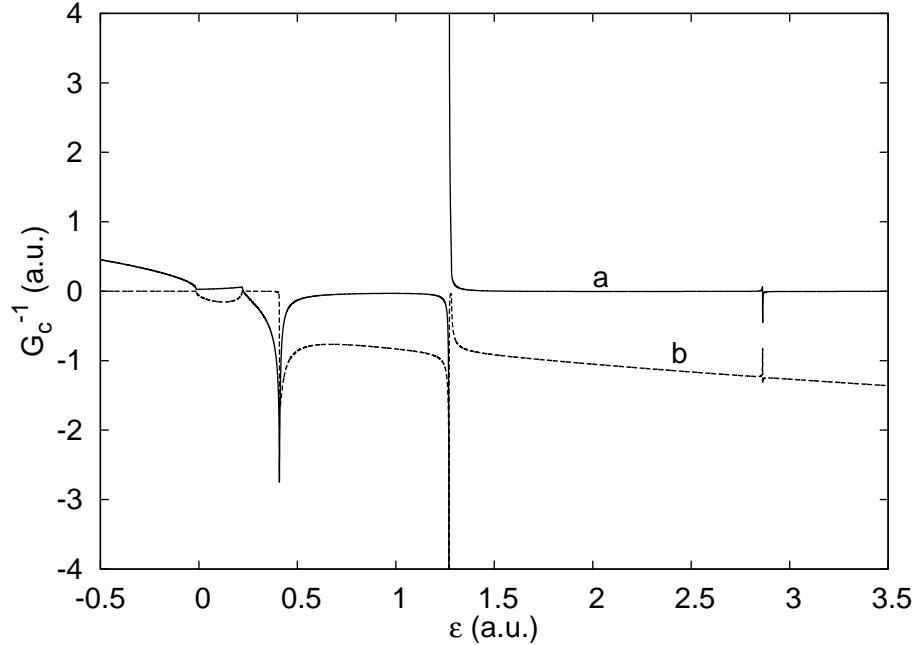}
\caption{Embedding potential for Cu(111) evaluated at $z_c=-10$ a.u., as a function of $\mathrm{Re}\,\epsilon$.
$\mathrm{Im}\,\epsilon$ is kept fixed at $2.5\times 10^{-4}$ a.u.
Curves \emph{a} and \emph{b}, real and imaginary parts of $G_c^{-1}(\epsilon)$.} \label{fig:cu_emb}
\end{center}
\end{figure}
Figure \ref{fig:cu_emb} shows the embedding potential calculated in this way for the Cu substrate in the [111] direction, 
evaluated at $z_c=-10$ a.u., with an imaginary part in the energy of $2.5\times 10^{-4}$ a.u. The singularities are at the band edges.

The next stage in finding $\bar{G}_c^{-1}(t)$ is to take the Fourier transform of
$G_c^{-1}(\epsilon)$, with the extra factor of $-1/\mathrm{i}\epsilon$ (\ref{eq:mod_FT}). 
At large $|\epsilon|$, the embedding potential has free-electron behaviour, 
proportional to $|\epsilon|^{1/2}$, and we can
ensure rapid convergence of the integral by subtracting this off and adding on the corresponding time-dependent potential, 
\begin{equation}
\bar{G}_c^{-1}(t)=\frac{\mathrm{i}}{2\pi}\int_{-\infty}^{+\infty}\!\!d\epsilon\;\frac{\exp(-\mathrm{i}\epsilon t)}{\epsilon}
[G_c^{-1}(\epsilon)-G_f^{-1}(\epsilon)]+\bar{G}_f^{-1}(t),  \label{eq:embpot2}
\end{equation}
where the free-electron embedding potentials are given by \cite{PhysRevA.56.763,Ehrhardt:1999vn,Inglesfield:2008ys}
\begin{equation}
G_f^{-1}(\epsilon)=\left\{\begin{array}{ll}\sqrt{-\epsilon/2}&\epsilon<0\\-\mathrm{i}\sqrt{\epsilon/2}&\epsilon>0\end{array}\right.
\!,\;\;\;
\bar{G}_f^{-1}(t)=\left\{\begin{array}{ll}0&t<0\\\frac{1-\mathrm{i}}{2\sqrt{\pi t}}&t>0\end{array}\right. . \label{eq:fe_embpot}
\end{equation}
We evaluate the integral in (\ref{eq:embpot2}) by discretization between finite energy limits; we smear out the band edge singularities 
in the integrand by taking $\epsilon$ with a small positive imaginary part (as in figure \ref{fig:cu_emb}), 
except in $\exp(-\mathrm{i}\epsilon t)$. 

\begin{figure}[h]
\begin{center}
\includegraphics[width=14cm] {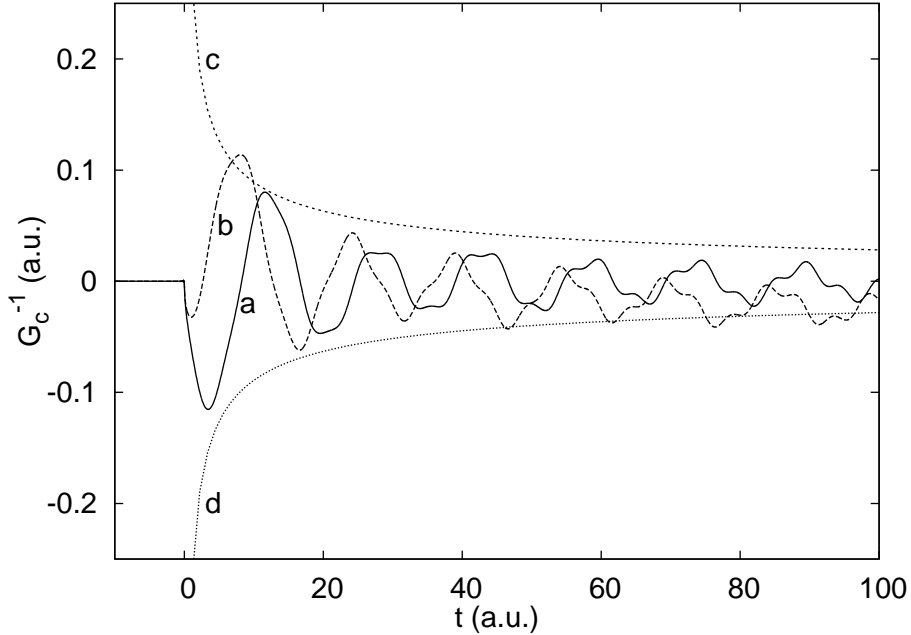}
\caption{Time-dependent embedding potential for Cu(111) evaluated at $z_c=-10$ a.u. 
Curves \emph{a} and \emph{b}, real and imaginary parts of $[\bar{G}_c^{-1}(t)-\bar{G}_f^{-1}(t)]$; 
curves \emph{c} and \emph{d}, real and imaginary parts of $\bar{G}_f^{-1}(t)$.} \label{fig:cu_emb_pot}
\end{center}
\end{figure}
Figure \ref{fig:cu_emb_pot} shows the the time-dependent 
embedding potential for Cu(111) $\bar{G}_c^{-1}(t)$, with the free-electron term 
subtracted; the free-electron embedding potential $\bar{G}_f^{-1}(t)$, with its inverse square root singularity
as $t\rightarrow 0$, is also plotted. The embedding plane is taken at $z_c=-10$ a.u. (figure \ref{fig:cupot}). 
In the evaluation of the integral, the energy is discretized with $\Delta\epsilon=1.25\times 10^{-4}$ a.u. 
between energy limits of $\pm 50$ a.u., with
an energy broadening of $\mathrm{Im}\,\epsilon=2.5\times 10^{-4}$ a.u. The accuracy of this procedure is shown
by the embedding potential satisfying causality very accurately, 
with $\bar{G}_c^{-1}(t)$ almost exactly zero for $t<0$. There are in fact some tiny Gibbs oscillations just below $t=0$ (figure \ref{fig:cu_emb_pot}). The structure above $t=0$ is interesting, showing several periodicities coming from the structure in $G_c^{-1}(\epsilon)$ (figure \ref{fig:cu_emb}). The fundamental period is 
$\Delta t\approx 15.2$ a.u., which in Fourier transform corresponds to an energy 
of 0.41 a.u. -- this is close to the energy 
difference of 0.42 a.u.  between the bottom of the band and the top of the first band gap. 
 
\begin{figure}[h]
\begin{center}
\includegraphics[width=12cm] {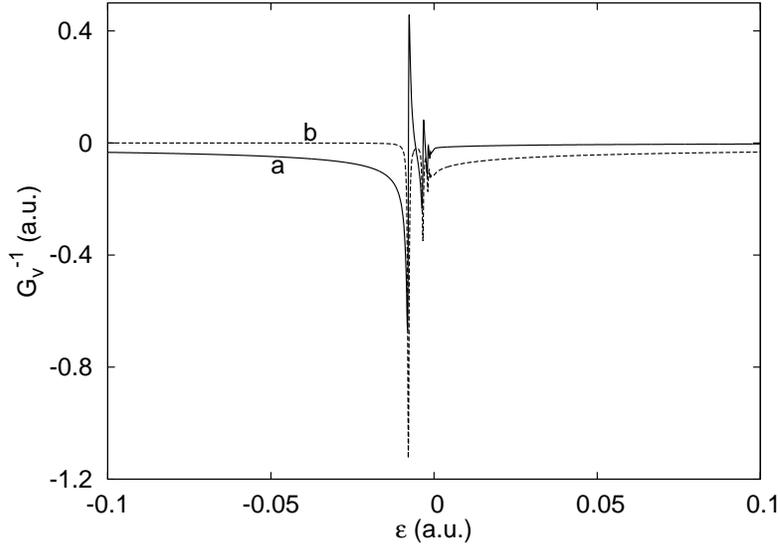}
\caption{Vacuum embedding potential evaluated at $z_v=10$ a.u., with $z_{\mathrm{im}}=2.1056$ a.u., as a function of $\mathrm{Re}\,\epsilon$
measured relative to the vacuum zero. $\mathrm{Im}\,\epsilon$ is kept fixed at $2.5\times 10^{-4}$ a.u.
Curves \emph{a} and \emph{b}, real and imaginary parts of $G_v^{-1}(\epsilon)$.} 
\label{fig:vac_emb}
\end{center}
\end{figure}

We now turn to the embedding potential to replace the vacuum region outside the crystal, in which asymptotically 
the electron feels the Coulomb tail of the image potential,
\begin{equation}
V(z)=-A_{10}-\frac{1}{4(z-z_{\mathrm{im}})}, \label{eq:image_pot}
\end{equation}
in terms of the Chulkov potential parameters (\ref{eq:chulkov}). We find the energy-dependent embedding potential from the logarithmic derivative expression (\ref{eq:en_logderiv}), which in one-dimension simplifies to
\begin{equation} 
G^{-1}_v(\epsilon)=-\frac{1}{2}\frac{\psi'(z_v)}{\psi(z_v)}, \label{eq:vac_deriv}
\end{equation}
where $\psi(z)$ is the outgoing or decaying solution of the Schr\"odinger equation. This is a combination of the regular and irregular
Coulomb functions $F_0$ and $G_0$ with angular momentum $L=0$, in the notation of Abramowitz and Stegun \cite{Abramowitz:1965sw}, 
\begin{equation}
\psi(z)=H^{-}_0(\eta,\rho)=G_0(\eta,\rho)-\mathrm{i}F_0(\eta,\rho) \label{eq:coulomb}
\end{equation}
with arguments given by
\begin{eqnarray}
\rho&=&\sqrt{2(\epsilon+A_{10})}(z_{\mathrm{im}}-z) \nonumber\\ 
\eta&=&\frac{1}{4\sqrt{2(\epsilon+A_{10})}}. \label{eq:arguments}
\end{eqnarray}
Thompson and Barnett \cite{Thompson:1986lo} give a rapidly converging continued fraction expression for $H^{-\prime}_0/H^{-}_0$,
from which we can immediately obtain $G^{-1}_v(\epsilon)$ using (\ref{eq:vac_deriv}). This is shown in figure \ref{fig:vac_emb}, evaluated
at $z_v=10$ a.u. with an image plane at $z_{\mathrm{im}}=2.1056$ a.u. In this figure we measure energy relative to the vacuum zero,
and the structure in $G^{-1}_v$ just below zero comes from bound states of the Coulomb potential.

To find the corresponding time-dependent embedding potential $\bar{G}_v^{-1}(t)$, we use the same procedure as in
(\ref{eq:embpot2}), subtracting the free-electron embedding potential inside the Fourier transform integral, and adding the time-dependent
free-electron embedding potential outside. 
\begin{figure}[h]
\begin{center}
\includegraphics[width=12cm] {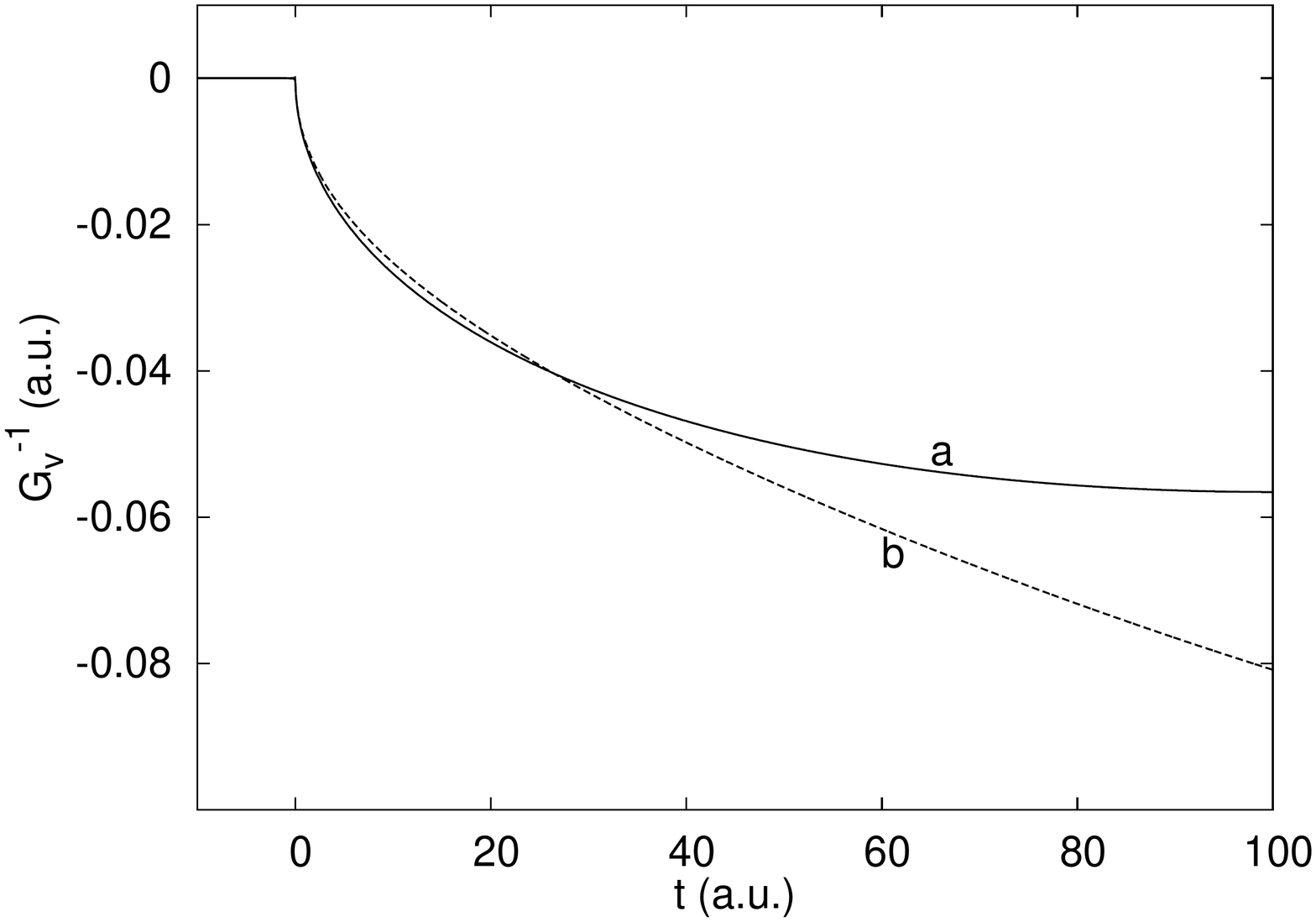}
\caption{Time-dependent vacuum embedding potential evaluated at $z_v=10$ a.u., with $z_{\mathrm{im}}=2.1056$ a.u. The image potential
is measured from the vacuum zero. Curves \emph{a} and \emph{b}, real and imaginary parts of
$[\bar{G}_v^{-1}(t)-\bar{G}_f^{-1}(t)]$.} \label{fig:image_time}
\end{center}
\end{figure}
The Fourier transform in the case of the image potential is more demanding computationally than for the crystal potential, 
and we use an energy discretization of $\Delta\epsilon=1\times 10^{-5}$ a.u. between energy limits of $\pm 50$ a.u., with an  
energy broadening of $\mathrm{Im}\,\epsilon=2.5\times 10^{-4}$ a.u. However,
even with this tiny value of $\Delta\epsilon$, $\bar{G}_v^{-1}(t)$ turns out to be finite, 
though small and constant, for $t<0$. This error, which is proportional to $\Delta\epsilon$, is actually constant over the whole range of $t$, positive as well as negative. This means that we can improve the 
accuracy -- essentially going to the limit of $\Delta\epsilon=0$ -- simply by subtracting 
off the constant value of $\bar{G}_v$ at negative $t$. 
The resulting time-dependent embedding potential is shown in
figure \ref{fig:image_time}. 
After the correction, the embedding potential is almost exactly zero for $t<0$, 
with even smaller Gibbs oscillations than in figure \ref{fig:cu_emb_pot}.
\section{Time-dependent tests} \label{sec:testing}
In this section we shall test the accuracy of the time-dependent embedding formalism, 
applying it to electron emission from the Cu(111) surface. 
But first we solve the time-independent Schr\"odinger equation to find 
the static electronic structure of the surface, using the Chulkov potential and the energy-dependent embedding
potentials $G_c^{-1}(\epsilon)$ and $G_v^{-1}(\epsilon)$ calculated in the last section. We determine the Green function 
$G(z,z';\epsilon)$ satisfying the embedded Schr\"odinger equation at complex energy $\epsilon$ with a small imaginary part, 
$\epsilon=E+i\eta$,
\begin{eqnarray}
\lefteqn{-\frac{1}{2}\frac{\partial^2 G}{\partial z^2}+
(V-\epsilon)G(z,z';\epsilon)+\delta(z-z_c)\left[-\frac{1}{2}\frac{\partial G}
{\partial z}+G_c^{-1}(\epsilon)G(z_c,z';\epsilon)\right]}\hspace*{3cm}\nonumber\\&&
\hspace*{-2cm}+\delta(z-z_v)\left[\frac{1}{2}\frac{\partial G}
{\partial z}+G_v^{-1}(\epsilon)G(z_v,z';\epsilon)\right]=\delta(z-z'),\;\;z,z'\mbox{ in region I}.\label{eq:time-independent}
\end{eqnarray}
The local density of states -- the electron density at energy $E$ -- is then given by \cite{Inglesfield:1981vn}
\begin{equation}
\sigma(z,E)=\frac{1}{\pi}\mbox{Im}G(z,z;\epsilon). \label{eq:lds}
\end{equation}
$G(z,z';\epsilon)$ is expanded in terms of basis functions $\chi_i(z)$, 
\begin{equation}
G(z,z';\epsilon)=\sum_{i,j}G_{ij}(\epsilon)\chi_i(z)\chi_j(z'), \label{eq:green_expansion}
\end{equation}
and then (\ref{eq:time-independent}) becomes the matrix equation,
\begin{equation}
\sum_j\left[H_{ij}+\Sigma_{ij}(\epsilon)-\epsilon S_{ij}\right]G_{jk}=\delta_{ik}, \label{eq:ti_matrix}
\end{equation}
with the Hamiltonian matrix given by
\begin{equation}
H_{ij}=\frac{1}{2}\int_{z_c}^{z_v}dz\frac{d\chi_i}{dz}\frac{d\chi_j}{dz}+\int_{z_c}^{z_v}dz\chi_i(z)V(z)\chi_j(z), \label{eq:ti_ham}
\end{equation}
the embedding matrix by
\begin{equation}
\Sigma_{ij}(\epsilon)=G_c^{-1}(\epsilon)\chi_i(z_c)\chi_j(z_c)+G_v^{-1}(\epsilon)\chi_i(z_v)\chi_j(z_v),  \label{eq:ti_emb}
\end{equation}
and the overlap matrix by
\begin{equation}
S_{ij}=\int_{z_c}^{z_v}dz\chi_i(z)\chi_j(z). \label{eq:ti_overlap}
\end{equation}
We use trigonometric basis functions,
\begin{equation}
\chi_m(z)=\left\{\begin{array}{l}\cos\frac{m\pi\zeta}{2D},\,
\;m\mbox{ even}\\\sin\frac{m\pi\zeta}{2D},\;\;m\mbox{ odd}\end{array}
\right., \label{eq:basis}
\end{equation}
where $\zeta$ is measured from the mid-point of region I,
\begin{equation}
\zeta=z-\frac{z_c+z_v}{2},\label{eq:mid-point}
\end{equation}
and a value of $D>(z_c-z_v)/2$ to give a range of logarithmic derivatives at
$z_c$ and $z_v$ for matching on to the embedding potentials. In the time-dependent applications
we construct an orthonormal basis from (\ref{eq:basis}), with the eigenvectors of the overlap matrix as coefficients.

\begin{figure}[h]
\begin{center}
\includegraphics[width=14cm] {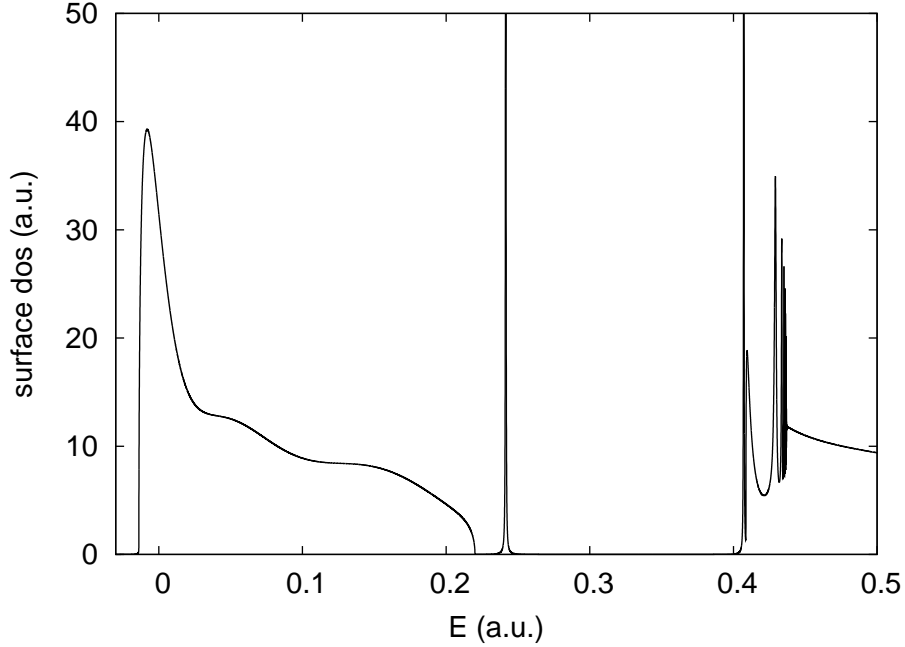}
\caption{Surface density of states of Cu(111), the local density of states integrated through region I, taken between $z_c=-10$ a.u. 
and $z_v=+10$ a.u. Energy broadening is $1\times 10^{-5}$ a.u.}\label{fig:surface_dos}
\end{center}
\end{figure}

Integrating the local density of states (\ref{eq:lds}) over the embedded region, we obtain the surface density of
states for Cu(111) shown in figure~\ref{fig:surface_dos}. In this calculation we take region I between $z_c=-10$ a.u. and $z_v=+10$ a.u., 
and use 40 basis functions, defined with $D=12$ a.u. in (\ref{eq:basis}); the imaginary part of the energy, 
which broadens the discrete states, is $1\times 10^{-5}$ a.u. These results, 
which correspond to zero wave-vector parallel to the surface in three dimensions, 
show the Shockley surface state at $E=0.2415$ a.u., close to the bottom of the band gap at $0.2201$ a.u, 
with the first image state at $E=0.4072$ a.u., immediately below the top of the gap
at $0.4087$ a.u. \cite{Jurczyszyn:1991fk}; the higher members of the image series overlap the continuum and become
resonances. For comparison, the Fermi energy is at $0.2556$ a.u.\cite{Chulkov:1999fh} and the
vacuum zero at $0.4371$ a.u. (table \ref{table:chulkov}). The surface state energies are almost
identical to the values given by Chulkov \emph{et al.} \cite{Chulkov:1999fh}, and close to experimental results \cite{PhysRevLett.50.526,Kubiak:1988kx}. A caveat is that 
although the Chulkov potential reproduces the electronic states near the s-p band gap, in 
particular the Shockley and image surface states, it does not describe the d-states, which lie
1 - 3 eV below the band gap \cite{Fluchtmann:1998fk}, nor does it give the band width 
accurately. 

We now return to solving the time-dependent embedded Schr\"odinger equation (\ref{eq:matvar2}) for Cu(111),
with a surface perturbation given by
\begin{equation}
\delta V(z,t)=A \exp-(z^2/\Xi)\sin(\omega t). \label{eq:pert}
\end{equation}
This perturbation is switched on at $t=0$, and we follow the subsequent time-evolution of $\phi(z,t)$ (\ref{eq:tgr}), 
with expansion coefficients $a_i$, given by equation (\ref{eq:matvar2}). The unperturbed wavefunction 
$\xi(z)$ (\ref{eq:eig},\ref{eq:tle}), which enters the time-evolution equation through vector $e_i$ (\ref{eq:inhom1}),  can be efficiently
calculated using the Numerov method \cite{Thijssen:1999xi}, integrating inwards from the vacuum.
The matrix elements of the Hamiltonian $\bar{H}_{ij}(t)$ (\ref{eq:hammat2}) are given by (\ref{eq:ti_ham}), with $\delta V(z,t)$ included,
\begin{equation}
\bar{H}_{ij}(t)=\frac{1}{2}\int_{z_c}^{z_v}dz\frac{d\chi_i}{dz}\frac{d\chi_j}{dz}+\int_{z_c}^{z_v}dz\chi_i(z)[V(z)+\delta V(z,t)]\chi_j(z),
\label{eq:td_ham}
\end{equation}
and the embedding matrix (\ref{eq:embmat}) has the same form as (\ref{eq:ti_emb}),
\begin{equation}
\bar{\Sigma}_{ij}(t)=\bar{G}_c^{-1}(t)\chi_i(z_c)\chi_j(z_c)+\bar{G}_v^{-1}(t)\chi_i(z_v)\chi_j(z_v). \label{eq:td_emb}
\end{equation}

To develop a simple time evolution algorithm, we write equation (\ref{eq:matvar2}) in abstract
notation,
\begin{equation}
\frac{da}{dt}=-\mathrm{i}[\bar{H}a+\Gamma],\;\;\mbox{where }\Gamma=e+\int_0^tdt'\bar{\Sigma}(t-t')\frac{da}{dt'}. \label{eq:cn1}
\end{equation}
Then to first order in the time interval $\delta t$,and using central differences, we have
\begin{equation}
a(t+\delta t)=a(t)-\mathrm{i}\delta t\bar{H}(t+\delta t/2)a(t)-\mathrm{i}\delta t\Gamma(t+\delta t/2). \label{eq:cn3}
\end{equation}
But to first order we can also write
\begin{equation}
a(t+\delta t)=a(t)-\mathrm{i}\delta t\bar{H}(t+\delta t/2)a(t+\delta t)-\mathrm{i}\delta t\Gamma(t+\delta t/2), \label{eq:cn4}
\end{equation}
and adding (\ref{eq:cn3}) and (\ref{eq:cn4}) gives
\begin{equation}
a(t+\delta t)=\left[1+\mathrm{i}\frac{\delta t}{2}\bar{H}\!\left(t+\frac{\delta t}{2}\right)\right]^{-1}
\left\{\left[1-\mathrm{i}\frac{\delta t}{2}\bar{H}\!\left(t+\frac{\delta t}{2}\right)\right]a(t)
-\mathrm{i}\delta t\Gamma\!\left(t+\frac{\delta t}{2}\right)\right\}. \label{eq:cn5}
\end{equation}
This is the formula we use to advance the wavefunction expansion coefficients 
forward in time.

\begin{figure}[h]
\begin{center}
\includegraphics[width=14cm] {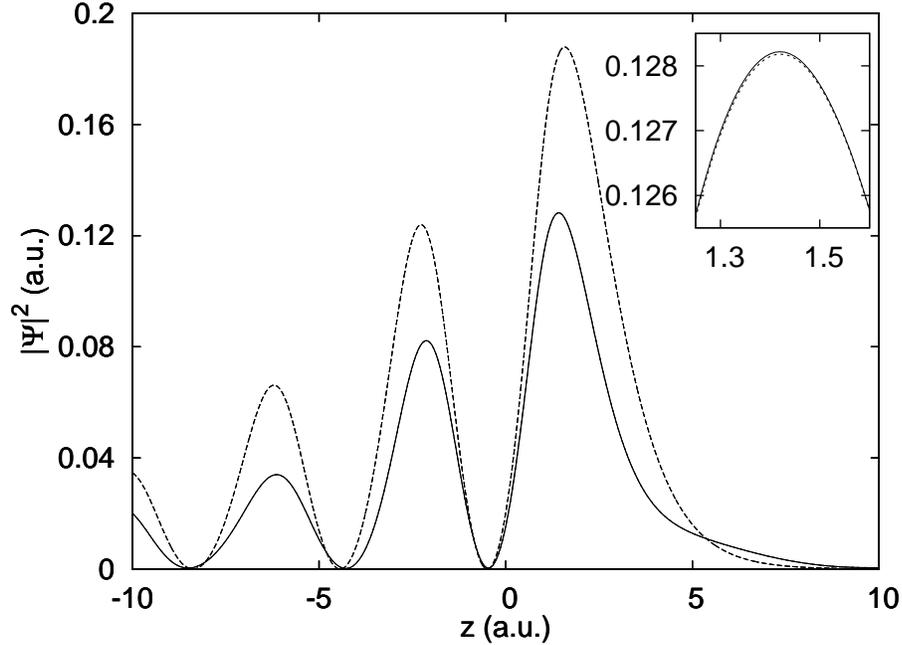}
\caption{Shockley surface state electron density at $t=200$ a.u. after applying the surface perturbation, calculated between $z_c=-10$ a.u. and $z_v=+10$ a.u.: solid line, 30 basis functions with $2D=22$ a.u. and $\delta t=0.016$ a.u.; short-dashed line, 40 basis functions with $2D=24$ a.u. and $\delta t=0.002$ a.u. Long-dashed line shows the unperturbed density.}
\label{fig:convergence}
\end{center}
\end{figure}

Although (\ref{eq:cn5}) turns out to be very accurate -- much more so than the time evolution method we described in our
earlier paper \cite{Inglesfield:2008ys} -- it is only stable for small $\delta t$. Because the trigonometric functions in (\ref{eq:basis})
are defined with respect to $2D$, but only used within the smaller embedding range, there is the risk of overcompleteness,
and the closer we are to linear dependency in the basis set, the smaller the time interval $\delta t$ needed for stability. Taking an embedding
range $(z_c-z_v)$ of 20 a.u., and $2D=24$ a.u., $\delta t=0.002$ a.u. gives stability with up to 40 basis functions. With an
embedding range of 40 a.u. and $2D=44$ a.u. this time interval gives stability with up to 75 basis functions. Bringing $2D$ closer
to the embedding range helps matters, and we find that with $(z_c-z_v)=20$ a.u., $2D=22$ a.u., and 30
basis functions we can go up to $\delta t =0.016$ a.u. In fact this is very accurate: 
figure \ref{fig:convergence} shows the electron density
$|\Psi(z,t)|^2$ (\ref{eq:tgr}) of the Shockley surface state at $E=0.2415$ a.u. 
(figure \ref{fig:surface_dos}),  calculated with this latter basis set
 and the large time interval, at $t=200$ a.u. after the application of the surface perturbation (\ref{eq:pert}) with $A=0.2$ a.u.,
$\Xi=2$ a.u. and frequency $\omega=0.5$ a.u. We see that it is almost identical to the calculation with the same embedding range, 
$2D=24$ a.u., 40 basis functions, and $\delta t=0.002$ a.u.; the difference is only visible at the enlarged scale of the inset.
It is tempting to take $2D=(z_c-z_v)$ to eliminate the possibility of overcompleteness. 
However,  for a given size of basis set this is less accurate,
and it is much better to use a basis set with $2D>(z_c-z_v)$, giving a range of derivatives at the embedding surfaces.

To test the embedding method itself, we compare the time evolution of states calculated with different embedding surfaces.
First we consider the system with $z_c=-10$ a.u., $z_v=+10$ a.u., taking the basis set with $2D=24$ a.u. and 40 basis functions,
and a time interval of $\delta t=0.002$ a.u. In the second case region I extends deeper into the crystal, with $z_c=-20$ a.u., and
further into the vacuum, with $z_v=+20$ a.u.; we use a basis set with $2D=44$ a.u., 70 basis functions, and the same time interval
of $\delta t=0.002$ a.u. Let us consider the time evolution of the continuum state with 
$E=0.1$ a.u. (figure \ref{fig:surface_dos}), with the same surface perturbation as before, 
$A=0.2$ a.u., $\Xi=2$ a.u., and $\omega=0.5$ a.u. The results for the two calculations are
shown in figure \ref{fig:convergence},
\begin{figure}[h]
\begin{center}
\includegraphics[width=14cm] {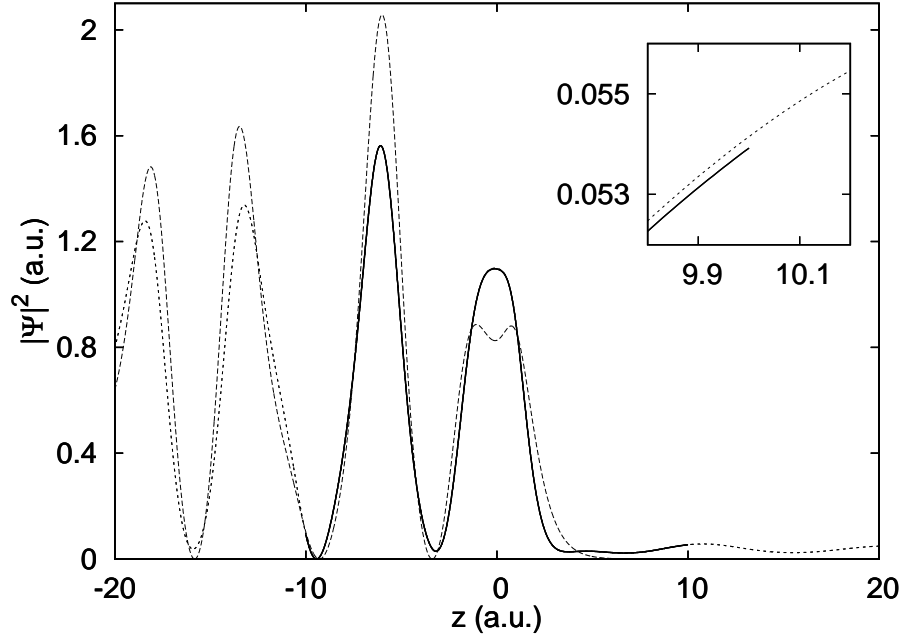}
\caption{Electron density of continuum state with $E=0.1$ a.u. at $t=200$ a.u. after applying the surface perturbation: solid line, $z_c=-10$ a.u., $z_v=+10$ a.u., 40 basis functions with $2D=24$ a.u.; 
short-dashed line, $z_c=-20$ a.u., $z_v=+20$ a.u., 70 basis functions with $2D=44$ a.u. In both cases time evolution proceeds with $\delta t=0.002$ a.u. Long-dashed line shows the
unperturbed density.}
\label{fig:emb_test}
\end{center}
\end{figure}
and once again the difference between the two curves is only visible in the magnified inset.

\begin{figure}[h]
\begin{center}
\includegraphics[width=13cm] {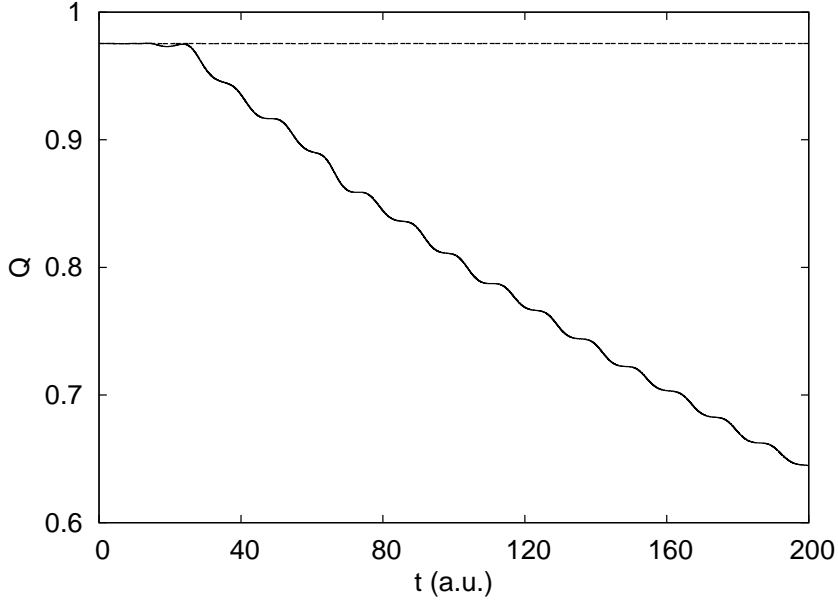}
\caption{Solid line, $Q(t)$, Shockley surface state electron number in the embedded region as a function of time; 
dashed line, $[Q(t)+J_c(t)+J_v(t)]$, electron number + time-integrated current across the embedding surfaces. 
$z_c=-20$ a.u., $z_v=+20$ a.u., 70 basis functions with $2D=44$ a.u.; time evolution proceeds with $\delta t=0.002$ a.u.}
\label{fig:charge_test}
\end{center}
\end{figure}
Another interesting test is to look at the number of electrons in the embedded region, $Q(t)$, as 
a function of time,
\begin{equation}
Q(t)=\int_{z_c}^{z_v}\!\!dz|\Psi(z,t)|^2.
\end{equation}
Continuity of charge requires
\begin{equation}
Q(t)+J_c(t)+J_v(t)=\mbox{constant}, \label{eq:continuity}
\end{equation}
where $J_c$ and $J_v$ are the time-integrated currents leaving region I across the left- and right-hand embedding surfaces,
\begin{equation}
J_{c/v}(t)=\int_0^t dt' j_{c/v}(t'),\;\;j_{c/v}=\mp\mbox{Im}\left.\Psi^*\frac{\partial\Psi}{\partial z}\right|_{z_c/z_v}.
\label{eq:current}
\end{equation}
We use the embedding formula (\ref{eq:dir_neu}) to calculate the normal derivative in the
expression for the current. Taking region I between $\pm 20$ a.u., 
70 basis functions with $2D=44$ a.u., and using $\delta t=0.002$ a.u., we calculate the time evolution of the surface state 
at $0.2415$ a.u. under the influence of the same surface perturbation as in the previous tests. The results are shown in
figure \ref{fig:charge_test}, and we see the steady decrease in the surface state charge
in the surface region as time progresses.  
However, the continuity equation (\ref{eq:continuity}) is satisfied remarkably accurately, 
with $[Q+J_c+J_v]$ staying almost perfectly constant -- the maximum variation is less than $10^{-4}$ across the whole time interval.
\section{Electron emission results} \label{sec:results}
Having tested the embedding method, we turn to the physics of electron emission from the Cu(111) surface. As in the previous section, 
we apply the surface perturbation given by (\ref{eq:pert}) -- this might correspond to the perturbation in surface photoemission with $p$-polarized light, when the
\textbf{E}-field of the incident light has a large component normal to the surface 
\cite{Levinson:1979ve}. 
Parameter $\Xi$ is taken as 2 a.u., and  we shall use amplitudes $A$ of 0.01 and 0.1 a.u. 
(for comparison, the amplitude of the
pseudopotential in the Chulkov potential for Cu(111) is 0.19 a.u., and the surface barrier is 0.44 a.u. (figure \ref{fig:cupot})). The results are  calculated using an embedded region
between $z_c=-20$ a.u. and $z_v=+20$ a.u., with 70 basis functions defined with $2D=44$ a.u.

\begin{figure}[h]
\begin{center}
\includegraphics[width=14cm] {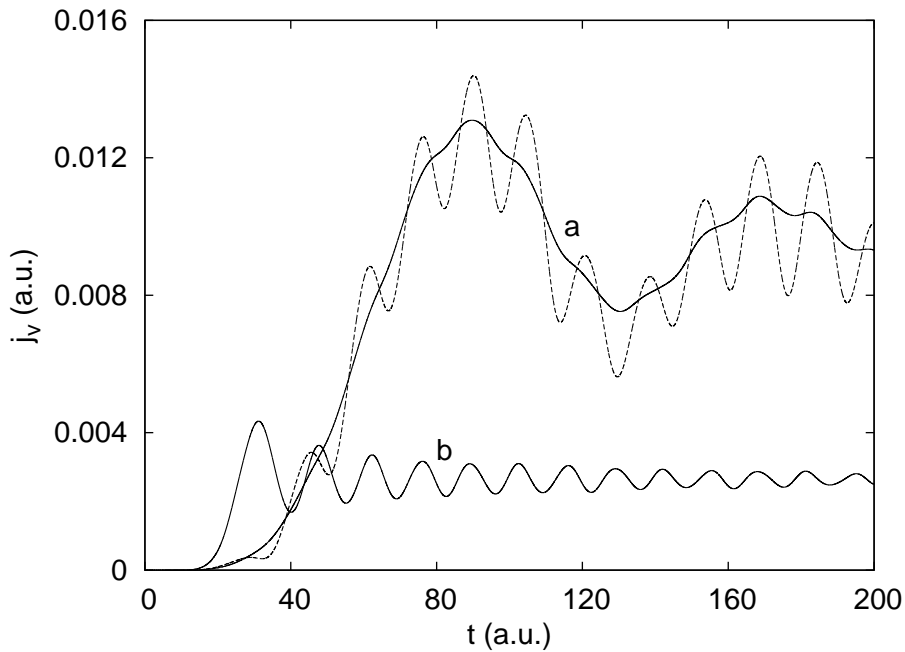}
\caption{Emission from the continuum state at $E=0.1$ a.u., showing current measured at 
$z_v=20$ a.u. as a function of time. Curves \emph{a}, 
$\omega=0.4$ a.u.: solid line, $j_v\times100$ with $A=0.01$ a.u.; dashed line, $j_v$ with $A=0.1$ a.u. Curve \emph{b}, 
$j_v\times 100$ with $\omega=0.8$ a.u., $A=0.01$ a.u.}
\label{fig:bulk_current}
\end{center}
\end{figure}
We begin with emission from the continuum state at $E=0.1$ a.u., normalized so that
the electron density equals the local density of states (\ref{eq:lds}), that is, energy normalization.
Figure \ref{fig:bulk_current} shows the current $j_v$ emitted into
the vacuum across the right-hand embedding plane at $z_v$ at two frequencies of perturbation, $\omega=0.4$ a.u. and 0.8 a.u.; 
we use amplitudes of $A=0.01$ a.u. and 0.1 a.u. at $\omega=0.4$ a.u., and 0.01 a.u. at $\omega=0.8$ a.u.
First the results for small amplitude, shown by solid lines in the figure: there is a
time-delay before the current reaches $z_v$ -- we shall discuss this later -- but after 
a large initial peak, the current shows oscillatory transients about a constant current. 
By curve-fitting (we fit a straight line to the time-integrated current $J_v$), we estimate 
that the average current after the initial peak is $1.0\times 10^{-4}$ a.u. for $\omega=0.4$
a.u. and $2.65\times 10^{-5}$ a.u. for $\omega=0.8$ a.u.  We can also calculate the long-term average current for small perturbations 
from Fermi's Golden Rule \cite{PhysRevB.10.4932,Hermeking:1975ys},
\begin{equation}
\bar{\jmath}_v=\frac{1}{k_f}|\langle i|\delta V|f\rangle|^2, \label{eq:fermi}
\end{equation}
where $|i\rangle$ is the initial state and $|f\rangle$ the final state, the well-known (though not particularly obvious) time-reversed
LEED state. $k_f$ is the wave-vector of the final state (the incident free-electron wave in the LEED state is $\exp(-\mathrm{i}k_f z)$),
coming from the density of final states in one dimension, $1/(2\pi k_f)$ -- the $2\pi$ cancels with the prefactor of $2\pi$ 
in the Golden Rule \cite{Merzbacher:1998fk}. 
The Golden Rule expression then gives $\bar{\jmath}_v=9.62\times 10^{-5}$ a.u. for $\omega=0.4$ a.u., 
and $\bar{\jmath}_v=2.65\times 10^{-5}$ a.u. for $\omega=0.8$ a.u., both at $A=0.01$ a.u., in agreement with the
time-dependent calculations. The decrease in current with increasing frequency is due to a combination of density of states 
and matrix element effects, as we can see from (\ref{eq:fermi}).

The lower the excitation frequency, the longer the period of the transients. Increasing the perturbation amplitude by a factor of 10 
at $\omega=0.4$ a.u., the current shows very pronounced oscillations of much shorter period on top of the long period oscillations
(dashed line, figure \ref{fig:bulk_current}). In fact even with $A=0.01$ a.u., these short-period oscillations are visible. 
Comparing the current for $A=0.01$ a.u. multiplied by a factor of 100 with that for $A=0.1$ a.u. (curves \emph{a}, figure \ref{fig:bulk_current}), 
we see that the amplitude of the short-period oscillations is non-linear in the intensity of the perturbation,
varying as the \emph{cube} of the perturbation amplitude $A$. 
This shows that these  oscillations
are the result of interference between first-order and second-order excitations from the bulk state; the frequency of the short-period
oscillations is exactly $0.4$ a.u., as we would expect for an interference term. This higher order term in the current is much less 
pronounced in emission with $\omega=0.8$ a.u.

\begin{figure}[h]
\begin{center}
\includegraphics[width=14cm] {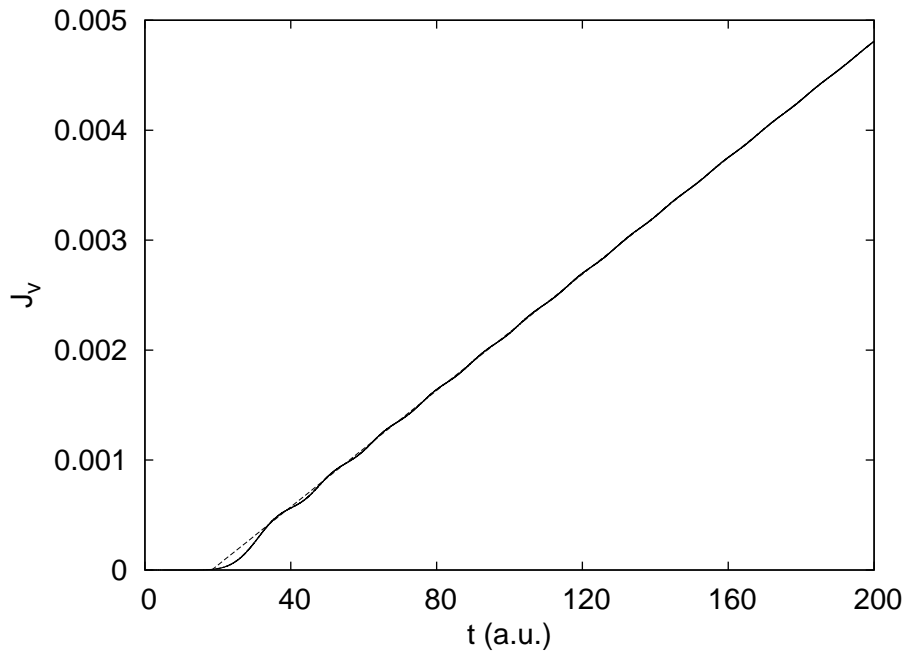}
\caption{Time-integrated current at $z_v=20$ a.u. emitted from the continuum state at $E=0.1$ a.u.,
$\omega=0.8$ a.u., $A=0.01$ a.u. Solid line, $J_v(t)$; dashed line, linear fit beyond $t=80$ a.u.}
\label{fig:fit_current}
\end{center}
\end{figure}

\begin{figure}[h]
\begin{center}
\includegraphics[width=14cm] {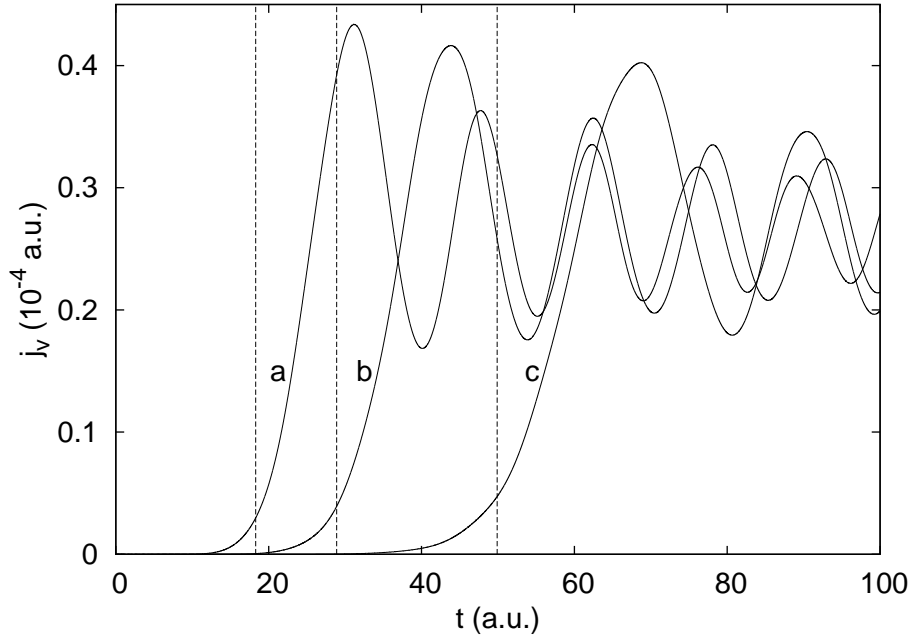}
\caption{Current emitted from the continuum state at $E=0.1$ a.u. measured at different planes as a
function of time, $\omega=0.8$ a.u., $A=0.01$ a.u. Curve \emph{a}, $z_v=20$ a.u.; \emph{b}, $z_v=30$
a.u.; \emph{c}, $z_v=50$ a.u.  The vertical lines show the effective arrival times $t_v$ from fits to the
time-integrated current.}
\label{fig:current_dispersion}
\end{center}
\end{figure}

From the curves for $j_v(t)$ shown in figure \ref{fig:bulk_current} it is clear that 
there is no unique time at which the electrons start to arrive at $z_v$, where we measure the current --
do we mean the time at which the current differs from zero, or perhaps when the first peak arrives? 
The first is ill-defined, as the increase from zero is gradual, and numerical limitations mean that we 
cannot determine it accurately. The second is too late, especially for the emission curves at $\omega=0.4$ a.u., 
with a final state energy just above the vacuum zero. However, we can define a convenient measure from the 
straight-line fit to $J_v(t)$, which we used to find the average current. Figure \ref{fig:fit_current} shows
the time-integrated current $J_v(t)$ evaluated at the right-hand embedding plane in emission from the 
continuum state with $E=0.1$ a.u., excited by the surface perturbation with frequency $\omega=0.8$ a.u. 
and amplitude $A=0.01$ a.u., fitted by a straight line beyond $t=80$ a.u. We define the effective arrival time 
$t_v$ for the current as the intercept of the linear fit with the time axis -- in this case, $t_v=18.3$ a.u. 
This compares with the classical arrival time, evaluated from the velocity of the electrons in the final state, 
of $20.8$ a.u. Figure \ref{fig:current_dispersion} shows $j_v(t)$ measured at different values of $z_v$, 
together with the arrival times $t_v$ determined from the linear fits: we see that our effective arrival
times lie close to the start of the steep initial rise in current. The classical arrival times are about 2 a.u. later, 
still not long after the start of the current rise, and well before the first peak. The time-structure of $j_v(t)$, with
its transients, presumably reflects the way that the perturbation (\ref{eq:pert}) is switched on suddenly at $t=0$:
a spectrum of energies is excited until the system settles down. This is why the initial rise in current widens 
when $j_v(t)$ is measured at larger $z_v$ (figure \ref{fig:current_dispersion}) -- $j_v$ disperses.

\begin{figure}[h]
\begin{center}
\includegraphics[width=14cm] {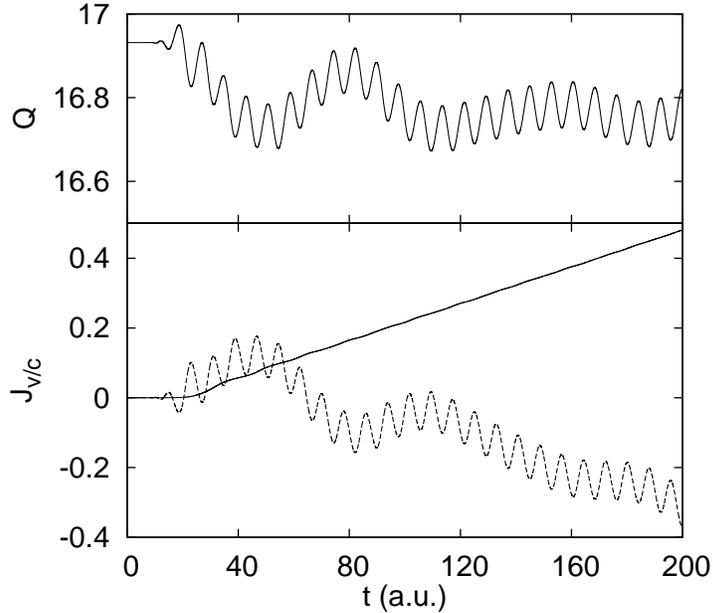}
\caption{Upper figure, electron number $Q(t)$ in the embedded region in emission from the continuum state 
at $E=0.1$ a.u., with $\omega=0.8$ a.u., $A=0.1$ a.u. Lower figure, time-integrated currents
across the embedding surfaces: solid line, $J_v(t)$; dashed line, $ J_c(t)$.}
\label{fig:bulk_int_current}
\end{center}
\end{figure}
Our most interesting results concern the electron number $Q(t)$ in the embedded
region, and how this is related to the time-integrated currents across the bulk and vacuum embedding surfaces -- we have already studied $[Q(t)+J_c(t)+J_v(t)]$  in section \ref{sec:testing}.
Figure \ref{fig:bulk_int_current} shows $Q(t)$, $J_c(t)$, and $J_v(t)$ for excitation
from the continuum state with energy $E=0.1$ a.u. by the surface perturbation with $\omega=0.8$
a.u. and amplitude $A=0.1$ a.u. (we use a larger amplitude to emphasize the results). Both
$Q(t)$ and $J_c(t)$ show continuing short-period oscillations, 
but if we average these by eye we see that after a decaying transient, $Q$ settles down 
to a fairly constant value. This is the result of the current leaving the surface 
region into the vacuum across $z_v$ being balanced by the current approaching the surface 
from the bulk across $z_c$ -- the slopes of $J_v(t)$ and $J_c(t)$ are almost equal and opposite
(remember, positive current means charge \emph{leaving} the surface region).
So in emission from a continuum state, the wavefunction is replenished from the bulk, and the 
electron number at the surface reaches a steady state. 

\begin{figure}[h]
\begin{center}
\includegraphics[width=14cm] {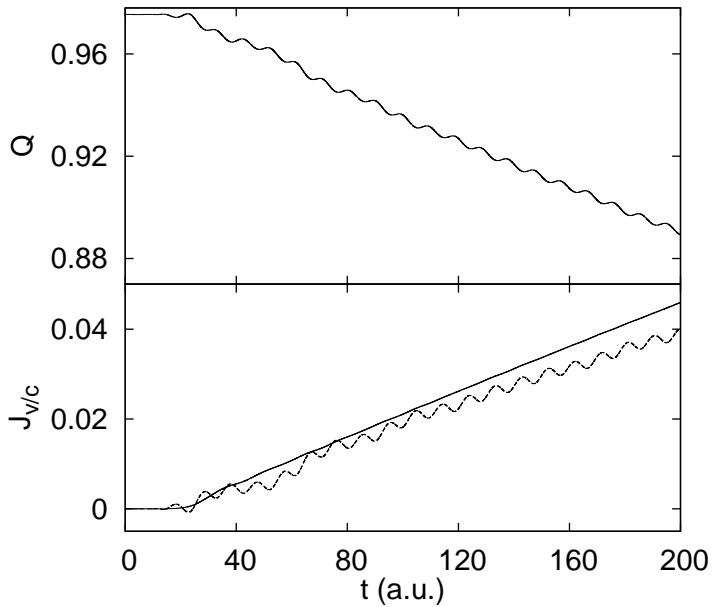}
\caption{Upper figure, electron number $Q(t)$ in the embedded region in emission from the surface state 
at $E=0.2415$ a.u., with $\omega=0.6585$ a.u., $A=0.1$ a.u. Lower figure, time-integrated currents across the embedding surfaces: solid line, $J_v(t)$; dashed line, $ J_c(t)$.}
\label{fig:surf_int_current}
\end{center}
\end{figure}
The behaviour is quite different in emission from the Shockley surface state at $E=0.2415$ a.u. 
Figure \ref{fig:surf_int_current} shows the number of electrons in the surface region and time-integrated
currents, with $\omega=0.6585$ a.u. (giving the same final state energy as in figure \ref{fig:bulk_int_current}) and $A=0.1$ a.u. 
We see that $Q$ drops steadily, as the surface
state is depopulated by currents leaving the surface into the bulk as well as into the vacuum: 
$J_c(t)$ is roughly equal to $J_v(t)$, both directed away from the surface. The currents correspond to electron momentum, so the near-equality
of $J_c(t)$ and $J_v(t)$ approximately conserves momentum 
(the surface step also provides a source
of momentum \cite{Levinson:1979ve}). On the other hand, in emission from the continuum state
momentum is directed from the bulk into the vacuum. 

One of the pleasing features of our work is that the embedding method describes the 
different time-evolution of the surface state and the continuum state, within the one-electron picture we are using. 
But we must remember that in real life, a state is depopulated in one go
by the emission of a single electron, and many-body effects repopulate the resulting hole. 
The finite mean-free path of the electrons also means that the bulk reservoir will not repopulate
the continuum state as efficiently as our calculation suggests. The difference between a localized
surface state and an extended continuum state may be more apparent than real after all.

\section{Outlook} \label{sec:outlook}
Having demonstrated and applied the method, we are aware of its limitations. 
Apart from the fact that it neglects many-body effects, there are restrictions to the
embedding method itself. Chief of these is 
the fact that the time for evolving the wavefunction through $\delta t$ is proportional to $t$, 
because the embedding term in the Hamiltonian involves an integral with the upper limit $t$ 
(\ref{eq:matvar2}). It is also quite time-consuming to evaluate the time-dependent embedding 
potential for the image potential region, though we can certainly go well beyond the present time-limit of 200 a.u. (about 5 fs), even with a desk-top computer. To reach much 
longer times further developments of the method are necessary.
In the meantime there is much to do with the existing surface programs, considering excitation by 
ultra-short light pulses, multi-photon processes and the like. Our time-dependent embedding method should have other applications, for example to adsorbates \cite{Achilli:2009uq} and metallic contacts \cite{Ishida:2008xd}, where time-independent embedding has proved useful.

\begin{acknowledgments}
I would like to thank S. Crampin and M. I. Trioni for their help.
\end{acknowledgments}
\bibliography{TD-1.bib}
\end{document}